\documentclass[twocolumn]{aastex63}
\usepackage{amsmath}
\usepackage{amssymb}
\usepackage{graphicx}
\usepackage{color}
\usepackage{CJKutf8}            % Chinese name
\usepackage{float}

\definecolor{darkgreen}{rgb}{0,0.35,0}
\shorttitle{Eccentric Planets, Wide Gaps}
\shortauthors{Chen et al.}

\newcommand\ypedit{\bgroup\markoverwith{\textcolor[rgb]{0.0, 0.5, .0}{\rule[0.5ex]{8pt}{1.5pt}}}\ULon}

\begin{document}

\title{Wide Dust Gaps in Protoplanetary Disks Induced by Eccentric Planets: A Mass-Eccentricity Degeneracy}

\author{Yi-Xian Chen\begin{CJK*}{UTF8}{gbsn}(陈逸贤)\end{CJK*}}

\affiliation{Department of Astrophysical Sciences, Princeton University, USA}

\affiliation{Department of Physics, Tsinghua University, Beijing, 100084 China} 
\author{Zhuoxiao Wang\begin{CJK*}{UTF8}{gbsn}(王卓骁)\end{CJK*}}
\affiliation{Institute for Advanced Studies, Tsinghua University, Beijing 100084, China}

\author{Ya-Ping Li\begin{CJK*}{UTF8}{gbsn}(李亚平)\end{CJK*}}
\affiliation{Theoretical Division, Los Alamos National Laboratory, Los Alamos, NM 87545, USA}

\author{Cl\'{e}ment Baruteau}
\affiliation{IRAP, Universit\'{e} de Toulouse, CNRS, UPS, Toulouse, France}

\author{Douglas N. C. Lin\begin{CJK*}{UTF8}{gbsn}(林潮)\end{CJK*}}
\affiliation{Department of Astronomy \& Astrophysics, University of California, Santa Cruz, CA 95064, USA}
\affiliation{Institute for Advanced Studies, Tsinghua University, Beijing 100084, China}

\correspondingauthor{Yi-Xian Chen}
\email{yc9993@princeton.edu}
\begin{abstract}
The tidal perturbation of embedded protoplanets {on} their 
natal disks has been widely attributed to be the cause of gap-ring structures in sub-mm images of 
protoplanetary disks around T Tauri stars. Numerical simulations of this process have 
been used to propose scalings of characteristic dust gap width/gap-ring distance
 with respect to planet mass. Applying such scalings to analyze observed gap samples yields a continuous mass distribution for 
a rich population of hypothetical planets in the range of several Earth to Jupiter masses.  In contrast, the
conventional core-accretion scenario of planet formation predicts a bi-modal mass function due to 1) the
onset of runaway gas accretion above $\sim$20 Earth masses and 2) suppression of accretion induced by
gap opening.  Here we examine the dust disk response to the tidal perturbation of eccentric planets 
as a possible resolution of this paradox. Based on simulated gas and dust distributions, 
we show the gap-ring separation of Neptune-mass planets with small 
eccentricities might become comparable to that induced by Saturn-mass planets on circular orbits. This degeneracy
may obliterate the discrepancy between the theoretical bi-modal mass distribution and the observed
continuous gap width distribution. Despite damping due to planet-disk interaction, modest eccentricity
may be sustained either in the outer regions of relatively thick disks or through resonant excitation
among multiple super Earths.  Moreover, the ring-like dust distribution induced by planets with small 
eccentricities is axisymmetric even in low viscosity environments, consistent with the paucity of vortices in ALMA images.
\end{abstract}


\section{Introduction}
\label{intro}

ALMA observations reveal an abundance of protoplanetary disks (PPDs) with cavity/gap and ring features in mm or sub-mm wavelengths \citep[e.g.][]{long2018,andrews2018,Huang2018,lodato2019,Long2019}. 
Typically, a depleted gap appears to separate an inner disk/ring with an outer bright ring,
a feature which could be explained by the perturbation of gas and dust profile by an embedded planet \citep{Goldreich_Tremaine_1980,             Lin_Papaloizou1986a, Lin_Papaloizou_1993,  paardekooper2006,Zhuetal2011,zhang2018}. 
The characteristic gap width
or the dust ring's distance from gap center can be retrieved from the brightness profiles, 
and applied to infer the mass of the planets and derive exoplanet demographics \citep{lodato2019,Nayakshin2019}. 

The scaling of the characteristic gap width/gap-ring distance $w$ as a function of planet mass and disk parameters has been 
extensively studied. This observable feature is regulated by 1) the excitation, 2) propagation, and 3) dissipation
of gas density waves excited at the planets' Lindblad and corotation resonances \citep{goldreichtremaine1979, Papaloizou1984,
Lin_Papaloizou1986a}. 
These waves carry angular momentum and their deposition in the disk changes the local surface density profile 
\citep{shu1985, Takeuchi1996}.
Based on the simulations of gas surface density  response to the tidal perturbation of isolated planets, \citet{DuffelChiang2015,Duffell2020} fitted the width in the \textit{gas gap} profile and found it to be dependent on the aspect ratio $h_0$ at planet location, associated with gas pressure. 
On the other hand, 
ALMA maps measure the continuum brightness of the disk surface, which characterizes the spatial distribution of the mm-size {\it dust 
particles} which are coupled to the gas to a limited extent. 
Under the influence of hydrodynamic drag, these grains diffuse amongst the disk gas.   Based on numerical simulations of the mm-size dust profile, 
\citet{rosotti2016, zhang2018, facchini2018} identified that the characteristic gap width/ring distance in dust emission is nearly independent of $h_0$. 
Despite this discrepancy, both \citet{Duffell2020} and \citet{zhang2018} indicated that the characteristic gap width, either in gas or dust, has little dependence on the viscosity parameter $\alpha$, in contrast to the gap 
depth, which is a sensitive function of both $\alpha$ and $h_0$ \citep{Kanagawaetal2015}.

Dependence on disk parameters aside,
uncertainties in the
observable gap width's dependence
on planet mass {are} also crucial with regards to the derivation of planet mass. 
In disk simulations with sub-mm particles, it's usually indicated that the general dust gap 
width scales as $w \propto q^\beta$, where $q=M_p/M_*$ is planet mass ratio and $\beta \sim 0.3-0.5$ 
\citep{Ovelar2013,rosotti2016, zhang2018, facchini2018}. \citet{rosotti2016} associated this scaling to the planet's Hill radius $R_H$
and proposed $w\approx k R_H \propto q^{1/3}$, with $k\approx 7.5$ a magnitude factor which might be very slowly increasing over time. 
Based on this $w-q$ relation, \citet{lodato2019,Nayakshin2019} derived a continuous planet mass distribution from ALMA samples of potentially planet-induced gaps, and reported no sub-Saturn desert feature in contrast to population synthesis models based on the classical core-accretion
scenario \citep{Ida_Lin_2004, Mordasini2012}. To be specific, the core accretion theory estimates an relatively 
short runaway timescale for super Earths with atmospheric mass comparable to core 
mass \citep{Pollack_etal_1996}, consistent with a sub-Saturn desert in the mature planets' bi-modal mass 
distribution since such transitory stages during runaway could rarely be preserved. 
This apparent disparity between theory and observation either raises the concern for slowing down runaway 
accretion  (Ali-Dib et al, submitted) or invalidates the core accretion scenario. 

In this paper, we consider an alternative possibility that the inferred mass from the observed gap data might be misinterpreted. 
While the validity of a universal {power-law index} of $\beta$ remains to be further examined, and consideration a few (20-30) samples may be misled  by intrinsic observational errors due to beam smearing, we draw attention to another hitherto neglected facet 
that may introduce much 
uncertainty in these inferences: the embedded planets' eccentricity, a possible result of planet–disk interactions \citep{Papaloizou2001,GS2003,KleyDirksen2006,lichen2021} when there is a cavity in the gas \citep{2021MNRAS.500.1621D} or interaction between multiple planet embryos \citep{Zhang2014}. The Hill radius scaling 
is inferred from simulations of planets on circular orbits, while planets with eccentric orbits might extend the characteristic 
width of the gap when their radial excursion is larger than their Hill radius \citep{Li2019,2021A&A...651A..90F}. 
In particular, planets with masses comparable to Neptune and non-negligible eccentricities  
{might} be mistakenly identified as Saturn-mass 
planets on circular orbits if their mass is extrapolated from the conventional approaches.
Such a mis-representation would lead to an
%gap width/gap-ring distance inference method, leading to 
overestimation on the occurrence rate for Saturns or sub-Saturns. 
Another motivation to investigate eccentric planets in ALMA disks is that they are known to induce shallower gaps than circular cases 
\citep[e.g.][]{DuffelChiang2015,Li2019}, which disfavors the theoretical prediction on 
the excitation of Rossby Wave Instability (RWI) and the 
generation of vortices \citep{Lovelace1999,li2000,Li2001,Lietal2005,Li2020}.  This effect would 
reconcile the tension between generally-inferred low viscosity in {T Tauri} disks and the 
lack of observed asymmetries \citep{Marel2021,Hammer2021}.

This work is organized as follows: 
in \S \ref{numerical} we describe the numerical setup of our hydrodynamical simulations, 
and in \S \ref{results} we compare simulation results for planets on circular/eccentric orbits, 
as well as in high/low viscosity environments. We also construct a simple statistical model for 
evaluating how planet eccentricity smooths out the distribution of gap-ring distances.
In \S \ref{summary} we summarize our findings. 

\section{Numerical Setup}
\label{numerical}
We explore the dependence of characteristic dust gap width/ring distance on the planet and disk parameters with Dusty FARGO-ADSG \citep{Barutea2008,Baruteau2008b,Baruteau2016}, 
which is a modified version of the grid-based code FARGO \citep{Masset2000}. 
For our 2D PPD model, we choose a passively irradiated and locally isothermal disk, 
where the midplane temperature is $T \propto r^{-1/2}$ at distance to host star $r$, and the aspect ratio

\begin{equation}
    h=\dfrac{c_s}{v_K}=\dfrac {H} r=h_0 \left( {\dfrac{r}{a_p}} \right)^{1/4},
\end{equation}
where $c_s$ is the sound speed, 
$v_K$ is the Keplerian velocity, 
$H$ is the scale height and $a_p$ is the planet's semi-major axis. 
The gas surface density profile is initialized as

\begin{equation}
    \Sigma_g = \Sigma_{g,0} \left(\dfrac{r}{a_p}\right)^{-1},
    \label{gasdensityeqn}
\end{equation}
such that we have approximately a constant accretion rate \citep{Franketal1992},

\begin{equation}
    \dot{M}_{*} \sim 3 \pi \nu \Sigma=3 \pi \alpha h_{\mathrm{0}}^{2} a_{\mathrm{p}}^{2} \Sigma_{\mathrm{0}} \Omega_{\mathrm{p}},
\end{equation}
where viscosity $\nu=\alpha c_s H$, $\alpha$ is the dimensionless viscosity parameter \citep{ShakuraSunyaev1973} and $\Omega_{\mathrm{p}}$ is the Keplerian angular velocity at $r=a_p$. The simulation domain ranges from $0.3 a_p$ to $3a_p$, with a grid resolution of $n_r \times n_\phi=512 \times 1024$ (logarithmic in the radial direction). We adopt $h_0=0.05$ and $\Sigma_{g,0} a_p^2=10^{-4}$. This corresponds to $\Sigma_{g,0}=8.9\text{\ g\ cm}^{-2}$ at $a_{\rm p}=10$au
%or $\Sigma_0=35.6\text{\ g\ cm}^{-3}$ at $a_{\rm p}=5$au 
for a central stellar mass of $M_{\odot}$. The initial dust-to-gas mass ratio $\Sigma_{d,0}/\Sigma_{g,0} = 0.01$ is set to be uniform and the dust is modelled as a pressure-less fluid. 

The dust species has an internal density of $\rho_\bullet=1.3\text{\ g\ cm}^{-3}$ and a fixed size of $s=0.675$mm, comparable to the ALMA observation wavelengths, in the Epstein regime where the Stokes number is given by, 
\begin{equation}
    \mathrm{St}=\frac{\pi s \rho_{\bullet}}{2 \Sigma_{\mathrm{g}}}=0.15 \times\left(\frac{s}{1 \mathrm{~mm}}\right)\left(\frac{\rho_{\bullet}}{1 \mathrm{~g} \mathrm{~cm}^{-3}}\right)\left(\frac{1 \mathrm{~g} \mathrm{~cm}^{-2}}{\Sigma_g}\right).
\end{equation}
For our adopted model parameters, {the dust's {initial $\mathrm{St}\approx0.015$ at $a_{\rm p}$}}.  For stars of different masses and different 
planet semi-major axis, the density of the disk as well as the dust size should be re-scaled accordingly. 

%respond to Ya-Ping (Zhang  et  al.  (2018)  showed  that performing simulations with dust of different size won’t significantly affect the w−q power scaling for circular planets,  albeit  introduce  small  changes  in  the  generalmagnitude. ) I have included this in the discussion section.
We neglect the dust 
feedback, but include dust diffusion with coefficient \citep{clarke1988,Zhuetal2011}:

\begin{equation}
    D=\nu \dfrac{1+4\mathrm{St}^2}{(1+\mathrm{St}^2)^2}\approx \nu
\end{equation}

At inner and outer boundaries, the gas velocities and densities are fixed to the initial steady-state value while the dust has an open inner boundary \citep{Meru2019,Sinclair2020}. We adopt the damping procedure \citep{deValborroetal2006} to avoid artificial wave reflection. We neglect the effect of accretion onto the planet and the smoothing length is $\epsilon=0.6H$.

Our model parameters are summarized in Table \ref{tab:para}. 
In the eccentric cases, we fix planet eccentricities at  $e=0.1$. 
%\yxedit{
The planet mass is initialized from $t=0$ as $M(t) = M_p \sin^2 (\pi t/2\tau_{damp})$ in the first 5 orbits ($\tau_{damp}=5 T_{orb}$) until it reaches the designated value. We run each simulation for 1000 orbits, 
equivalent to a timescale of $\sim$0.03 Myrs if the planet is located at 10au or $\sim$0.15 Myrs for 30au. %These timescales are a fraction of disk ages of the DSHARP sources.
%\yxedit{
Since the simulations are run for less than a viscous time, we neglect evolution of the inner boundary condition of the gas \citep{Dempseyetal2020}.

\begin{table}[htbp]
  \centering
  \begin{tabular}{lccc|c}
  
     \hline\hline
     model & $q$ & $e$ & $\alpha$ & $w'/r_{\rm gap}$ \\

     \hline
     nep\_a3 & 1e-4 & 0 & 1e-3 & 0.29\\
     nep\_a4 & 1e-4 & 0 & 1e-4 & 0.23\\
     nep\_e01a3 & 1e-4 & 0.10 & 1e-3 & 0.40 \\
     nep\_e01a4 & 1e-4 & 0.10 & 1e-4 & 0.33 \\
     
     \hline
     sat\_a3 & 3e-4 & 0 & 1e-3 & 0.37\\
     sat\_a4 & 3e-4 & 0 & 1e-4 & 0.27 \\
     sat\_e01a3 & 3e-4 & 0.10 & 1e-3 &  0.41 \\
     sat\_e01a4 & 3e-4 & 0.10 & 1e-4 & 0.34\\

     \hline\hline
   \end{tabular}
   %\end{center}
   \caption{
   %\yxedit{
   Fiducial model parameters: the planet mass ratio $q$, eccentricity $e$, viscous parameter $\alpha$, and normalized ring distance $w^{\prime}/r_{\rm gap}$, where $r_{\text {gap }}=\left(r_{\text {out}}+r_{\text {in}}\right) / 2$ and $w^{\prime}=r_{\text {peak}}-r_{\text {gap}}$, the gap boundaries $r_{\rm in}$, $r_{\rm out}$ is determined by where azimuthally-averaged dust density drops below a threshold value { ($f=0.1$)}, and $r_{\text {peak}}$ is location of the azimuthally-averaged dust density maxima. The planet semi-major axis is fixed to be $10$au where the initial gas surface density is $\Sigma_{g,0}=8.9\text{\ g\ cm}^{-2}$. See \S \ref{stnumber} for details on 8 additional runs.}
   %}
   \label{tab:para}
  \end{table}
  
\section{Results}
\label{results}
In Table~\ref{tab:para}, we provide an overview of our fiducial models and simulation results.

In \S \ref{intro}, we generally referred to the dust gap width or gap-ring separation as $w$. 
In previous investigations, these quantities differ slightly in definition although they are 
somewhat similar in magnitude. In our analysis of simulation results, we specifically 
consider the quantity $w^\prime$, defined as the gap-ring separation in the dust profile. 
The reasons are stated below.

{Based on the radial boundaries of the gap structure where the azimuthally-averaged dust density drops below a fraction $f$ of the initial density $\Sigma_{d,0}$, 
%\yxedit{we could define 
the gap center $r_{\rm gap}$ and a \textit{genuine} gap width $w$ could be defined as $r_{\rm gap}=(r_{\rm out}+r_{\rm in})/2$ and $w=(r_{\rm out}-r_{\rm in})/2$, analogous to the definition of \citet{DuffelChiang2015,Duffell2020} in gas density or \citet{rosotti2016} in brightness profile \footnote{We neglect the density spike induced by corotating dust around L4 and L5 points in the determination of the inner and outer boundaries}.
However, such a definition of $w$ may be sensitive to the threshold value $f$ when the gap is shallow. Also, \citet{rosotti2016} found that (for planets on circular orbits) the power slope of $w-q$ changes considerably over time (see their Fig 16). 
Alternatively, their results indicate the radial distance (or separation) from $r_{\rm gap}$ to the ring brightness peak $w'=r_{\rm peak}-r_{\rm gap}$ has a power dependence on $q$ that is relatively unchanged over $\sim2500$ orbital timescales, albeit there is a slight increase in the general magnitude (see their Fig 17). {Moreover, the characteristic widths of the gaps from \citet{long2018} and the DSHARP surveys are usually Gaussian dispersions fitted from the intensity profile, which is more determined by the entire realm of density fluctuation (minimum to peak), rather than comparison with threshold values.} Considering these factors, 
we choose $w'$ for our studies, and define $r_{\rm peak}$ as location of the dust density maxima in our case.
In cases with significant double-gap structures, we only consider the outermost primary gaps.}

\begin{figure}
  \centering
  \includegraphics[width=1.05\linewidth]{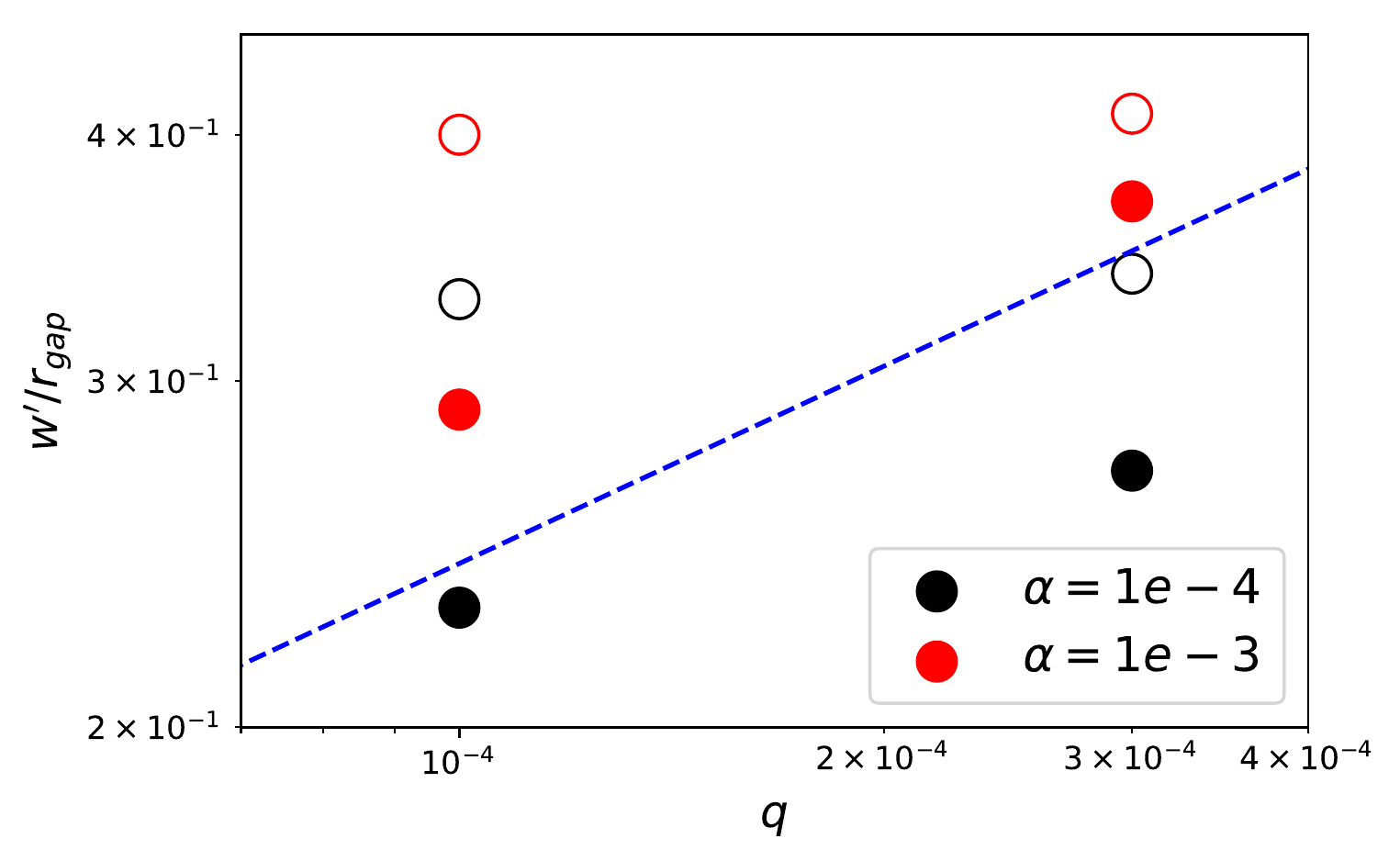}
  \caption{Summary of normalized ring distance $w'/r_{\rm gap}$ in all our cases, using $f=0.1$ to define the gap center. 
   The black signs indicate $\alpha=10^{-3}$ cases and the red signs indicate $\alpha=10^{-4}$ cases; circular cases are marked in filled circles and $e=0.1$ cases correspond to open circles. A scaling of $w/r_{\rm gap}=7.5R_H$ is plotted in blue dashed lines for reference.}
  \label{fig:summary}
\end{figure}

In the last column of Table \ref{tab:para}, we display the dimensionless dust gap width $w'/r_{\rm gap}$ for threshold value 
$f=0.1$ (see Fig. \ref{highvis1d}), we have also applied $f=0.5$ to determine $r_{\rm gap}$ and found no significant difference \footnote{{we normalized $w^\prime$ with $r_{\rm gap}$ following \citet{rosotti2016}, since the semi-major axis is not directly derivable from density and emission profiles.}}. 
In Fig \ref{fig:summary}, 
we summarize the $w^\prime/r_{\rm gap}$ in all 8 cases. 
The red signs indicate the high viscosity cases ($\alpha=10^{-3}$) and the black signs indicate the low viscosity cases ($\alpha=10^{-4}$); 
circular cases are marked in filled circles and $e=0.1$ cases are marked in open circles. 
We have also plotted 
an $\alpha$-independent approximate scaling {prescription  $w'/r_{\rm gap}\sim 7.5R_H/r_{\rm gap}\approx7.5(q/3)^{1/3}$ \citep{rosotti2016}   
in blue dashed lines for reference}. 

%\yp{I think this is the most important paragraph showing that the eccentric planets can open wide gaps compared to the circular ones. This should be emphasised since I don't find further discussion in other places. }

In the few samples presented here, 
we see that compared to circular cases, 
the normalized gap-ring distances in the $e=0.1$ cases are generally larger and has almost no dependence of planet mass. 
%\yxedit{In other words, 
These  results imply $w^\prime/r_{\rm gap}$ is nearly solely determined by $e$ when eccentricity is larger than a few times the Hill radius. 
This deduction is intuitive since with $e=0.1$, the radial excursion of the eccentric planet during one orbit is $0.2 a_p$, 
which is already much larger than the Hill radius and is comparable to the gap widths/gap-ring distances in circular cases, 
therefore we expect the eccentric orbits to push the gap edges and the pressure maxima farther out. 
For planets with masses {slightly larger than} Neptune ($q=10^{-4}$), 
such a subtle expansion is 
sufficient for the gap to become comparable in width to a gap carved out by a Saturn-mass planet
with a circular orbit. This expansion effect is smaller for Saturns ($q=3 \times 10^{-4}$) with 
a larger Hill radius, albeit eccentricity still widens a the gap-ring distance by $10-20\%$, until 
it reaches more or less the same value as the eccentric Neptune. {Our main intention is to point 
out that this widening produces a misleading effect, such that super Earths and Neptunes with 
high eccentricity might be mistakenly characterized as higher-mass planets on circular orbits, and 
leading to an overestimation of Saturns in the planet mass demographics.}

Although the general magnitude of $w^\prime/r_{\rm gap}$ has slightly increases with $\alpha$, similarly to the finding of \citet{zhang2018}
\footnote{{note that the gap width in \citet{zhang2018} is defined by the radial distance between inner and outer gap boundary, different from this work.}}, the offsets of our data points from the approximated viscosity-free Hill radius scaling is still small ($\lesssim 20\%$) in all circular cases. The difference may be attributed to the methods used to measure $w^\prime/r_{\rm gap}$
from the actual $\Sigma_{\rm d}$ profile rather than the brightness profile. This small discrepancy is comparable to the intrinsic resolution errors of 
the ALMA maps, which introduce as much ambiguities in the inference of 
the planets' mass from the gap width for individual systems \citep{Nayakshin2019}. Nevertheless, these offsets do not 
%\yxedit{destroy}
erase the apparent dependence of $w^\prime$ on $q$ $(w^\prime/r_{\rm gap} \propto R_H)$ for circular planets, 
%\yxedit{
and a more conspicuous influence of $e$ when $e \gg R_H/a_{\rm p}$. Uncertainties in the exact form of this somewhat degenerate $w^\prime-(q,e)$ relation can be minimized with a statistical sample 
in the inference of mass distribution from the characteristic gap-ring distance distribution, which we directly address in \S\ref{sec:widthdistribution}. Some specific comments on individual simulations are expressed in the following subsections.

\subsection{High Viscosity}
\label{sec:hivis}

\begin{figure}
  \centering
  \includegraphics[width=1.0\linewidth]{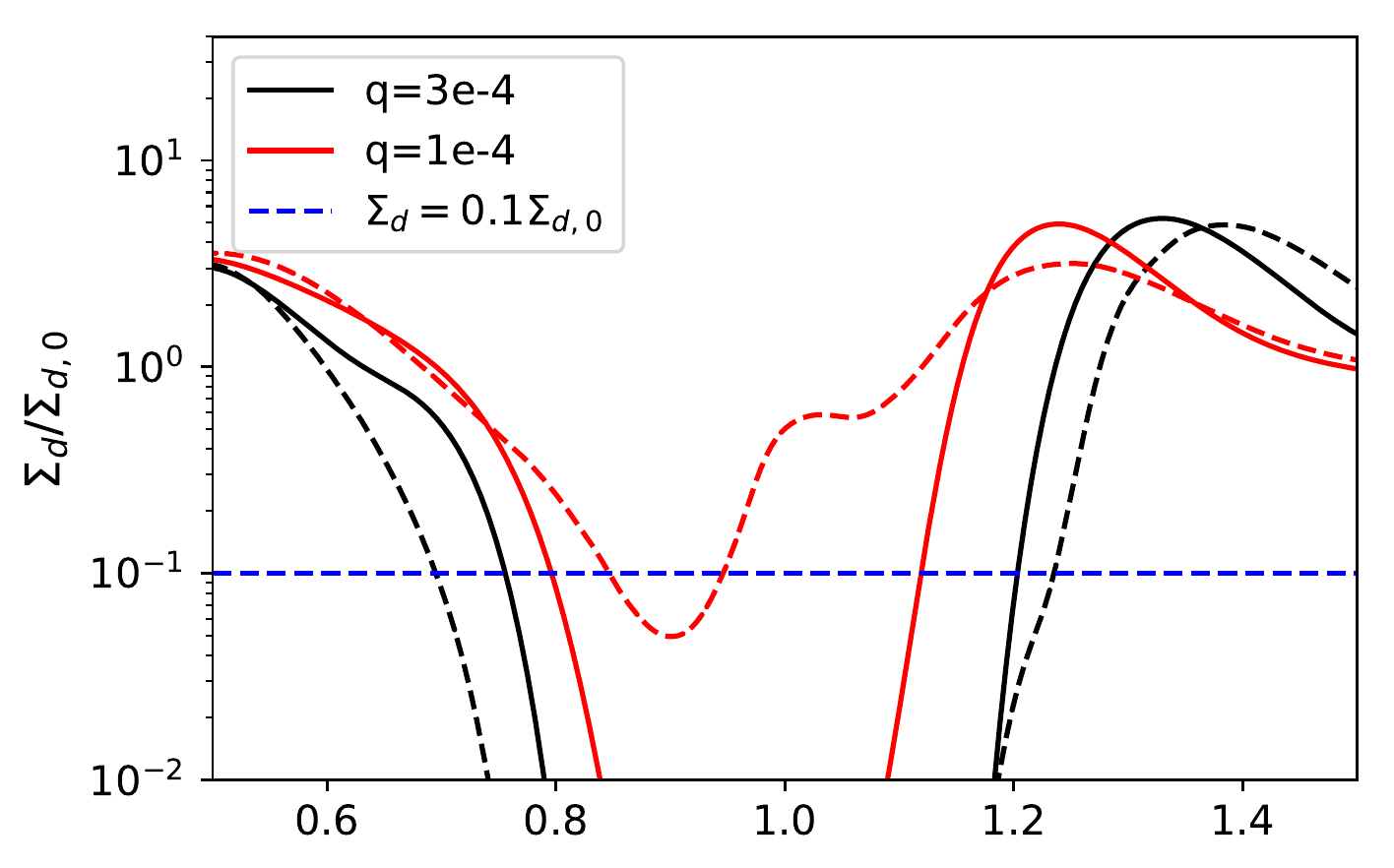}
  \caption{The azimuthally-averaged dust surface density in high viscosity ($\alpha=10^{-3}$) cases. The black lines indicate the Saturn cases ($q=10^{-4}$) and the red lines indicate the Neptune ($q=3\times 10^{-4}$). The dashed lines correspond to eccentric 
  { ($e=0.1$)} cases. The blue dashed line shows the threshold which we used to define the gap center $r_{\rm gap}$. }
  \label{highvis1d}
\end{figure}

We show the azimuthally-averaged dust profiles in Fig \ref{highvis1d} for the four cases with $\alpha = 10^{-3}$. The dust profiles are quite symmetric, similar to the findings of \citet{rosotti2016}.
The eccentric cases have shallower gaps than the circular cases, but the dust density maxima is pushed farther out from the gap center such that $w^\prime/r_{\rm gap}$ is larger. {In a somewhat outstanding case of the eccentric $q=10^{-4}$ planet, an expansion of $w'/r_{\rm gap}$ {compared to the circular cases} seems to be dominated by a shrink in $r_{\rm gap}$. 
This uncertainty may occur when the dust gap is shallow and highly asymmetric in the inner and outer radial directions, 
but we emphasize that in all other cases where eccentric planets can still open up a radially symmetric gap, $w'$ is still mainly associated with a retreat of the density maxima, especially in low viscosity cases.}

\begin{figure}
  \centering
  \includegraphics[width=1.0\linewidth]{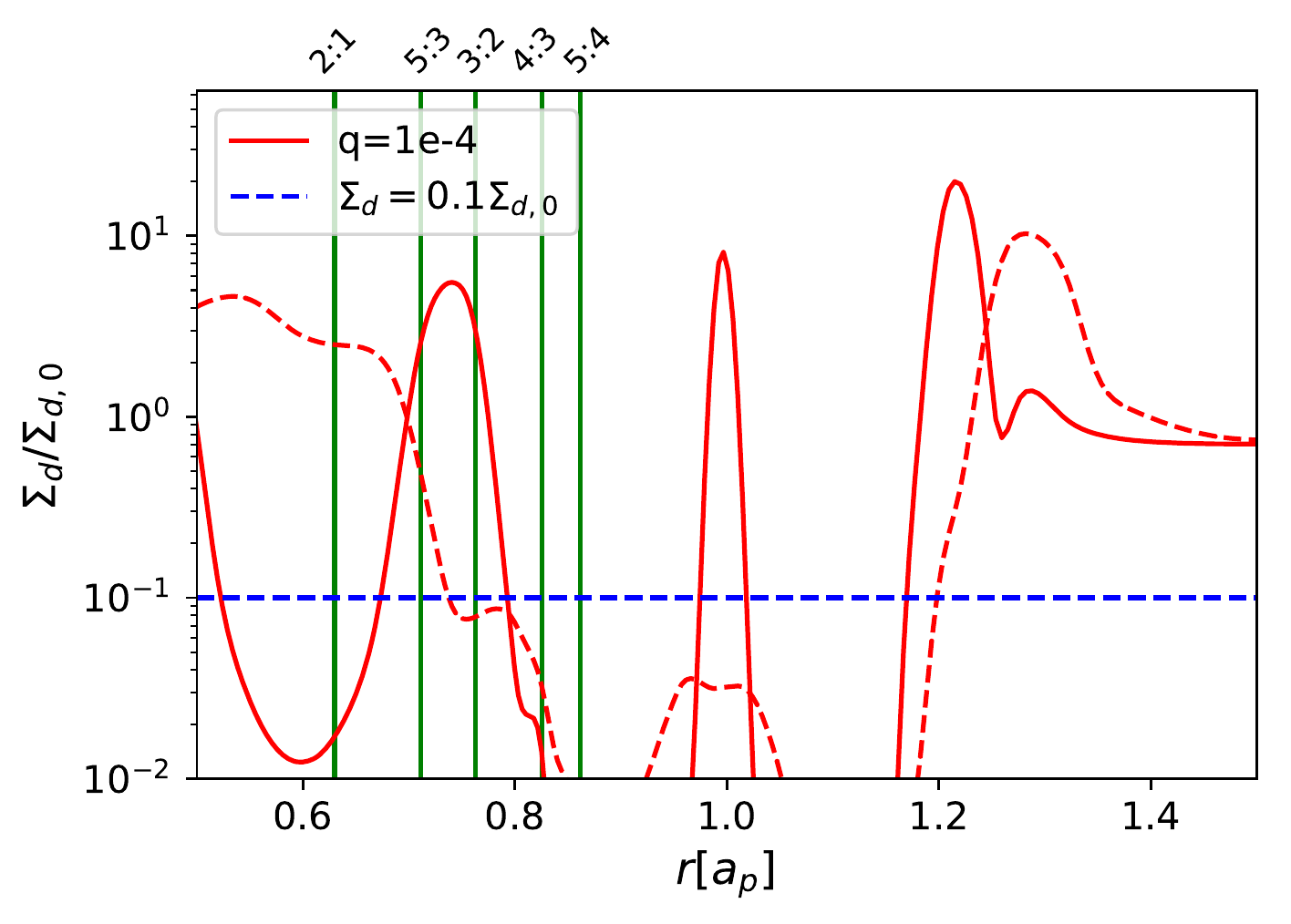}
  \includegraphics[width=1.0\linewidth]{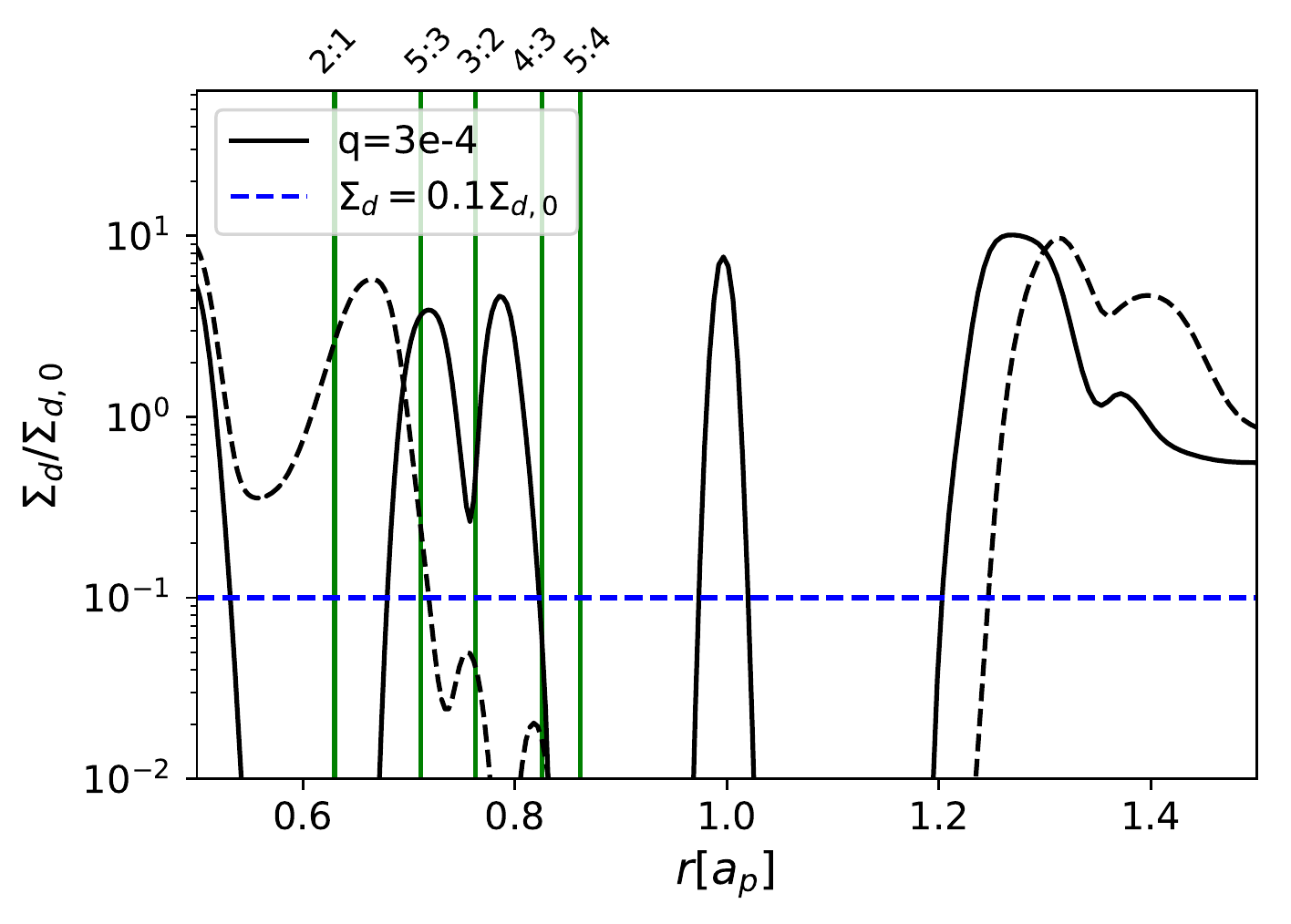}
  \caption{ The azimuthally-averaged dust surface density in low viscosity ($\alpha=10^{-4}$) cases. The legends are the same as Fig \ref{highvis1d}, we plotted the $q=10^{-4}$ and the $q=3\times 10^{-4}$ cases separately for clarification. The green vertical lines indicate the first-order {mean motion} resonances.}
  \label{low1d}
\end{figure}

\subsection{Low viscosity}
\label{sec:lowvis}
Although a moderate viscosity is in accordance with the high symmetry in ALMA observations, 
recent computational studies have shown that non-ideal MHD effects should suppress the MRI in the disk midplane \citep{bai2013,bai2017} where the temperature is not high enough to support a high ionization level \citep{Gammie1996}. 
By analyzing the ALMA CO line data, \citet{Flaherty2015,Flaherty2017,Flaherty2020} were able to put upper limits of turbulence in a number of planet-forming disks to be $\alpha \leq 10^{-3}$, 
suggesting a common low viscosity feature. Therefore, 
it is meaningful to probe the regime of lower viscosity.

\begin{figure*}
  \centering
  \includegraphics[width=1.0\textwidth]{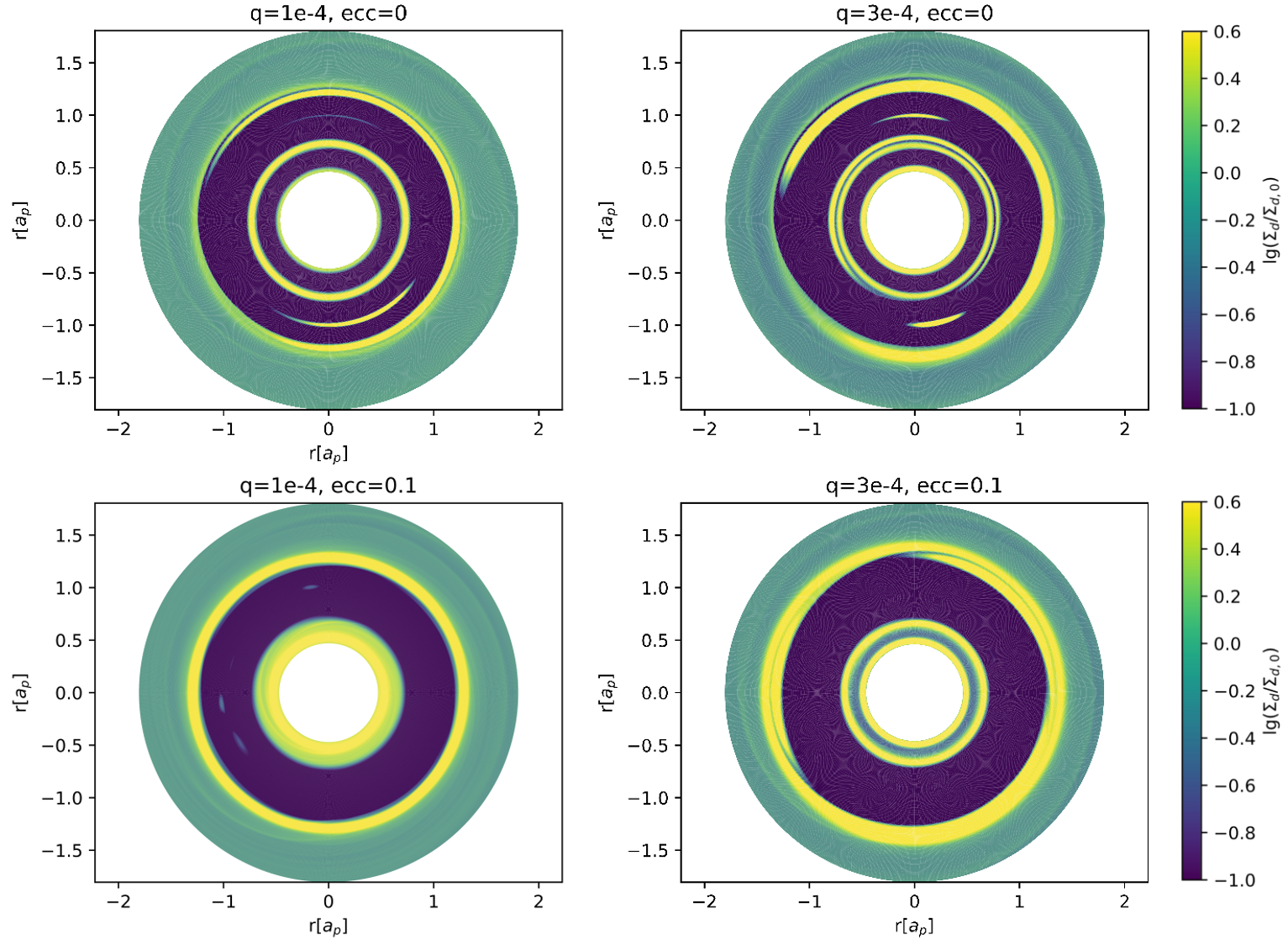}
  \caption{2D dust surface density distribution in the low viscosity cases. Colorbars denote the value of $\log_{10} (\Sigma_d/\Sigma_{d,0})$.}
  \label{2d}
\end{figure*}

While gap-ring features for our eccentric planets in the high viscosity regime might have low density contrast and avoid detection, the gaps are much deeper and the density contrasts much higher in the $\alpha=10^{-4}$ cases (see Fig \ref{low1d}). Nevertheless, the trend of gap-widening by eccentric planets is preserved. 
Another important feature of gaps carved out by circular planets in low viscosity environment is a secondary gap \citep{Dong2017,Dong2018,Huang2018}, 
possibly appearing at resonance locations.  Density waves are excited at 
the planets' Lindblad and corotation resonances, including their 2:1 mean motion resonances 
\citep{goldreichtremaine1979, Goldreich_Tremaine_1980}.  They carry angular momentum flux as they 
propagate through the disk \citep{Goldreich_Tremaine_1982, Papaloizou1984}. Viscous and nonlinear 
dissipation of these waves lead to angular deposition which changes the local $\Sigma_{\rm g}$ profile
\citep{shu1985, Lin_Papaloizou1986a, Takeuchi1996}.  This process introduces $\Sigma_{\rm g}$ variations
in the proximity of the low-order Lindblad resonances, which in turn modifies the propagation and nonlinear 
dissipation of the density waves.

%We suspect this difference may be due the active feedback in the disk response, especially near the Lindblad resonances where the density waves are launched. Moreover the difference in the $w$ versus $w^\prime$ definition may enlarge the dispersion in their values obtained by various teams.  

In eccentric cases, the planets 
also excite density waves at the eccentric resonances.  Although the torque density induced by a planet 
at its lowest eccentric resonances exerted on the disk gas is a factor of $e$ times weaker than those 
at the Lindblad resonances, their occupation in the regions between the Lindblad resonances leads to 
resonance overlap, modifying the launch, propagation, and dissipation of the density waves and 
smearing out secondary gap structures. Nevertheless, taking only the primary gap into calculation, 
we find the azimuthally averaged profiles give similar trends in $w'/r_{\rm gap}$ for comparison as the high viscosity case (see Fig \ref{fig:summary}).

It is also worth mentioning that we trace evidence of non-axisymmetry in the dust gaps opened by circular planets, similar to \citet{zhang2018}. 
This is demonstrated in the full 2D distribution of dust density in Fig \ref{2d}. The asymmetries/vortices in circular cases are due to two main reasons
a) the Rossby wave instability (RWI) at very steep gap edges where the gas vortensity gradient is high \citep[e.g.][]{Lietal2005} and 
b) retention of corotating dust particles around L4 and L5 points due to low diffusivity $ D \propto \alpha$. 
However, 
the vortices are smeared out in the eccentric cases, 
since an eccentric orbit not only quenches the accumulation of corotation materials (since L4 and L5 are no longer stationary points), 
but also prevents RWI by making gas gaps shallower and less steep. 
This is a potential solution to reconcile the general low viscosity in PPDs with the lack of observed symmetries in ALMA \citep{Hammer2021,Marel2021,Michel2021}, 
although this issue is not the main focus of our paper.
%\subsection{Saturn Cases}

\subsection{General Distribution of $w'/r_{\rm gap}$ under the Influence of Eccentric Planets: A Toy Model}
\label{sec:widthdistribution}

In \S\ref{sec:hivis} and \S \ref{sec:lowvis}, we have provided examples where eccentricity could disguise a Neptune-mass planet as a Saturn-mass 
planet in terms of gap-ring distance. Moreover, we suggest that the uncertainties associated with the calibration of the degenerate $w^\prime - (q,e)$ relationship may be
bypassed with the inference of a $q$ distribution from the observed $w^{\prime}/r_{\rm gap}$ distribution, based on the statistical 
analysis of comprehensive data. In this section, we directly address the problem of how a degenerate $w^\prime - (q,e)$ relationship might produce misleading features in the general gap-ring distance distribution.

Incorporating our previously discussed idea that the actual gap width/ring distance is determined by both the planet's gravitational
reach and the radial excursion, our small collection of samples in Fig \ref{fig:summary} imply a scaling of 

\begin{equation}
    w^{\prime}/r_{gap}\sim \max[AR_H/a_p, Be].
    \label{wrformula}
\end{equation}
%\yxedit{
While more extensive simulations are required to verify the variable's proportionality to $e$ in the limit of large eccentricity and constrain the scaling coefficients $A,B$, we adopt 
$A=7.5$ and $B=3.5$ based on the results of above simulations. {$B=3.5$ is modestly estimated from our $w^{\prime}/r_{\rm gap}$ results with $e=0.1$ and we neglect the small dependence on viscosity}. For the purpose
of illustration with a simple statistical analysis, this approximation is adequate.
We use a Monte Carlo approach to generate a distribution of $w'/r_{\rm gap}$ from planet mass and eccentricity distributions and compare results for different postulated eccentricity distributions.

Motivated by \citet{Zhou2007,Ida2013}, 
we apply a Rayleigh distribution to approximate the planets' eccentricity distribution:
\begin{equation}
    P(e) = \frac{e}{{\sigma_e}^{2}} \exp \left(-\frac{e^{2}}{2 {\sigma_e}^{2}}\right), \ \ \ \  e\in [0,1]
    \label{edis}
\end{equation}
where $\sigma_e$ denotes the characteristic dispersion as well as the eccentricity that corresponds to the peak probability. 
We also use a simple log-normal bi-modal distribution for planet mass: 
\begin{equation}
\begin{aligned}
    P(m) \propto& \dfrac{1}{\sqrt{2\pi} \sigma_m}  
    \exp{\dfrac{-(m-m_1)^2}{2\sigma_m^2}}\\
    +&\dfrac{1}{3\sqrt{2\pi} \sigma_m}
    \exp{\dfrac{-(m-m_2)^2}{2\sigma_m^2}}, 
    \ \  m \in[0.5, 3.5]
     \label{mdis}
     \end{aligned}
\end{equation}
where $m=\log_{10} (M_p/M_{\oplus}),m_1=1,m_2=2.5,\sigma_m=0.35$, the peak at Jupiter mass ($m_{2}$) is $1/3$ of the super Earth ($m_{1}$) peak value, 
motivated by the distributions from core accretion theory \citet{Ida_Lin_2004,Ida2013}. {Although generally more massive planets are more likely to excite and maintain eccentricity,
we neglect the correlation of $e$ and $m$ distribution in our simple analysis.}

\begin{figure}
  \centering
  \includegraphics[width=1.05\linewidth]{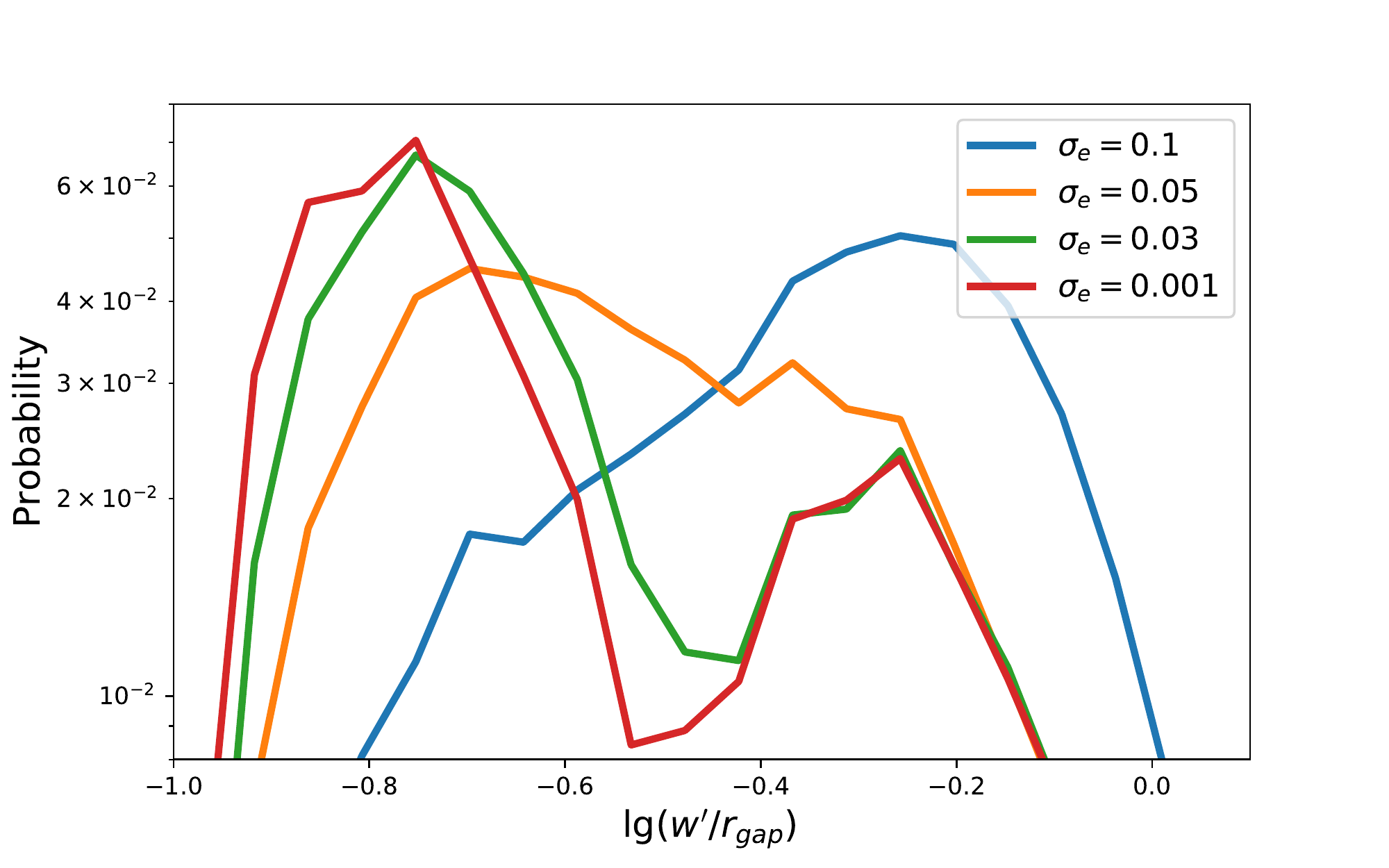}
  \caption{
%  \yxedit{
  The normalized distribution of $w'/r_{\rm gap}$ across 20 logarithmically distributed bins. The intrinsic mass distribution is assumed to be
  bi-modal (Eqn \ref{mdis}) and we assume a simple mass \& eccentricity dependent scaling for $w'/r_{\rm gap}$ (Eqn \ref{wrformula}). 
  Different colors correspond to different Rayleigh distributions of eccentricity (Eqn \ref{edis}), parameterized by the value $\sigma_e$.}
  \label{distribution}
\end{figure}

For four values of $\sigma_e$, we each generate $10^6$ planet mass-eccentricity configurations, 
and collect the results for $w'/r_{gap} 
\in [0.1-1.25]$ 
%responce to YP, in footnote
\footnote{{The possibility of eccentric/circular gas giants creating an observable gap-ring feature with $w' \gtrsim a_p$ 
is not completely ruled out, as in \citet{lodato2019,Nayakshin2019}.}} using Equation (\ref{wrformula}), categorizing the normalized occurrence 
rate into 20 logarithmically distributed bins. 
Although identification of gaps width/ring distance in the lower and upper end may suffer from heavy observational uncertainties, we are mostly concerned about the general shape of the distribution, 
which is insensitive to the specific range and binsize of catogorization. In Fig \ref{distribution}, 
the solid red line corresponds to a case  with nearly zero eccentricity,
which reflects the intrinsic bi-modal mass distribution. 
As we expand the eccentricity dispersion, while $\sigma_e = 0.03$ 
(green line) still preserves the general bi-modal shape, a moderate dispersion of $\sigma_e \gtrsim 0.05$ (orange line) is enough to smooth out
the double peak feature, 
since dust gaps induced by smaller planets have a much higher probability of being significantly widened 
by $e \gg R_H/a_p$. Comparing with Figure 2 of \citet{Nayakshin2019}, models with $\sigma_e \gtrsim 0.1$ (blue line) can be ruled out since they tend to eliminate all narrow gap/low mass planet
samples.  
This toy model provides a simple illustration on how planet eccentricity could bring uncertainties in the $w'/r_{\rm gap}$
distribution, 
which should be treated with caution when applied to infer planet mass. %\yp{Do you want to explain the disparity mentioned in the third paragraph of Introduction?}

\subsection{Dependence on Stokes Number}
\label{stnumber}

%\yxedit{
In the above simulations, only a single dust species is considered. 
In reality, dust population in PPDs is distributed across different sizes, and its coupling abilities with gas varies with Stokes number. In practice, ALMA observations trace brightness profiles \citep[e.g.][]{Ovelar2013,Pinilla2014}, which generally reflects the distribution of mm and sub-mm dust particles on the order of detection wavelengths \citep{Dong2017,Dong2018} which dominates the opacity contribution. Since potential planets' orbital radii may cover a large distance, the gas surface density in the these planets' vicinity also varies from case to case, and plays an important role in determining the $\mathrm{St}$ number of the characteristic dust species around observed ring structures. This issue has motivated us to perform 8 additional simulations with $a_p=30$au and $\Sigma_{g,0}=3\times 10^{-4}$. This setup corresponds to the same fiducial disk profile described by Eqn \ref{gasdensityeqn} at a larger distance, where $\Sigma_{g,0}=3 \mathrm{g/cm}^3$ is lower in physical units and the characteristic dust species' $\mathrm{St}$ is {raised to $
\approx 0.04$}. The simulation results are shown in Figure \ref{additional}, analogous to previous Figures \ref{fig:summary}, \ref{highvis1d} and \ref{low1d}. We briefly summarize our findings as follows:
%}

\begin{figure*}
  \centering
  \includegraphics[width=1.0\linewidth]{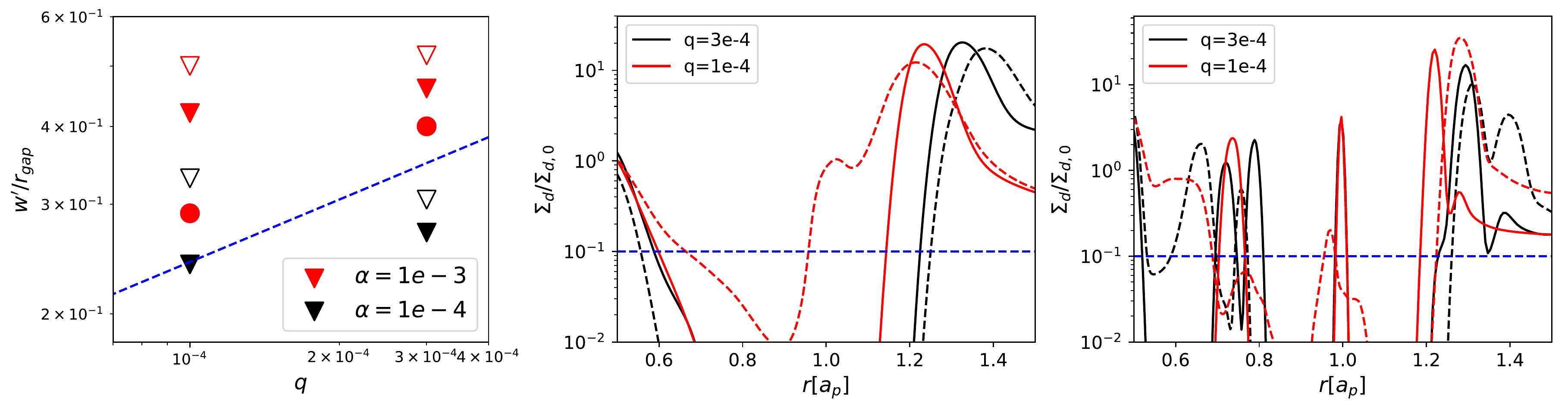}
  \caption{
  {Left panel: summary of all normalized ring distances in our additional cases ($a_p=30$au, dust St$\approx$0.04 for $\Sigma_{g,0}=3 \mathrm{g/cm}^3$) are plotted in wedge signs, analogous to Fig \ref{fig:summary}. The results for $q=10^{-4}, \alpha=10^{-3}$ cases in the fiducial runs ($a_p=10$au, dust St$\approx$0.015 for $\Sigma_{g,0}=8.9 \mathrm{g/cm}^3$) are plotted in red circles to indicate that a 3-fold increase in dust St alone may introduce a flattening of the $w^{\prime}-q$ scaling; Middle \& right panel: The azimuthally-averaged dust surface density in high \& low viscosity additional simulations. The legends are the same as Fig \ref{highvis1d} and \ref{low1d}. }}
  \label{additional}
\end{figure*}

%\yxedit{
First, comparing the solid and hollow wedge signs in the left panel of Figure \ref{additional}, we see that data points from our additional runs still demonstrate the effect of large eccentricity in smoothing out the $w^{\prime}/r_{gap} -q$ correlation in circular cases, that our general conclusion remains unchanged under small variation in the dust Stokes number. 
%}

%\yxedit{
Additionally, the general profiles and morphology of the dust distribution in our additional runs are similar compared to their corresponding fiducial cases (see middle and right panels of Fig \ref{additional}). We have also confirmed in the low viscosity cases, planet eccentricity smooths out the non-axisymmetric vortices existent in circular cases.
%}

%\yxedit{
Despite the general resemblance, dust gaps in all our additional cases are generally wider and the density contrasts are more 
pronounced. This difference arises because dust with larger $\mathrm{St}$ drifts faster over the same amount of time, leading 
to further retreat of the dust ring and severer depletion of the inner disk even when the general pressure gradient condition 
is similar \citep[][see their \S 5.2]{zhang2018}. However, we also note that this expansion is not linear with respect to 
$\mathrm{St}$, since by plotting two circular cases against our additional results in the left panel of \ref{additional}, 
we see that relative differences between circular cases of different $q$ are narrowed compared to fiducial cases, in other 
words, the $w^{\prime}-q$ scaling for circular cases tends to be flatter for larger Stokes number.
%}%\yp{Did you see any dependence on St for eccentric cases ? Both circular and eccentric cases have larger gaps}

%\yxedit{
In this comparison, $\mathrm{St}$ has only changed by a moderate factor of 3. Future simulations could expand the parameter space to include PPDs at different stages, where $\Sigma_{g,0}$, or equivalently the characteristic $\mathrm{St}$ of observed dust particles, can vary in $\sim$2 orders of magnitude. Nevertheless, we point out that extensive simulations in \citet{zhang2018} also implied that for dust-gas distributions with moderate maximum Stokes number of dust $\mathrm{St}_{max} \gtrsim 0.01$, the $w^{\prime}-q$ scaling reflected in the observed sub-mm images becomes shallower for larger $\mathrm{St}_{max}$, 
as gap widths for low-mass planets are significantly widened  \footnote{Note that in their Figure 14, the variation in $\mathrm{St}_{max}$ is controlled by both tuning $\Sigma_{g,0}$ and including different dust species in their synthetic image, (e.g. the green, red and purple solid lines corresponding to a same dust size distribution but ascending $\Sigma_{g,0}$; While the brown dashed line correspond to the same $\Sigma_{g,0}$ with the green case but different dust size distribution) and these two scenarios may deserve separate analysis.}.
This trend confirms what is qualitatively seen from our results, but a more rigorous $w^{\prime}-\mathrm{St}$ relation still appeals to further study, therefore for simplicity we did not include the dependence on $\mathrm{St}$ in our previous statistical analysis. On the other hand, we comment that if it's really the case that large $\mathrm{St}_{max}$ for observed dust shares a similar effect with high eccentricity on flattening out the $w^{\prime}-q$ profile, it would introduce yet another degeneracy into the statistical models and obscure the intrinsic bi-modal structure in gap width distribution even more.
%}

\subsection{Eccentricity Excitation and Damping}
\label{sec:eccentricitydamping}
\begin{figure}
  \centering
  \includegraphics[width=1.05\linewidth]{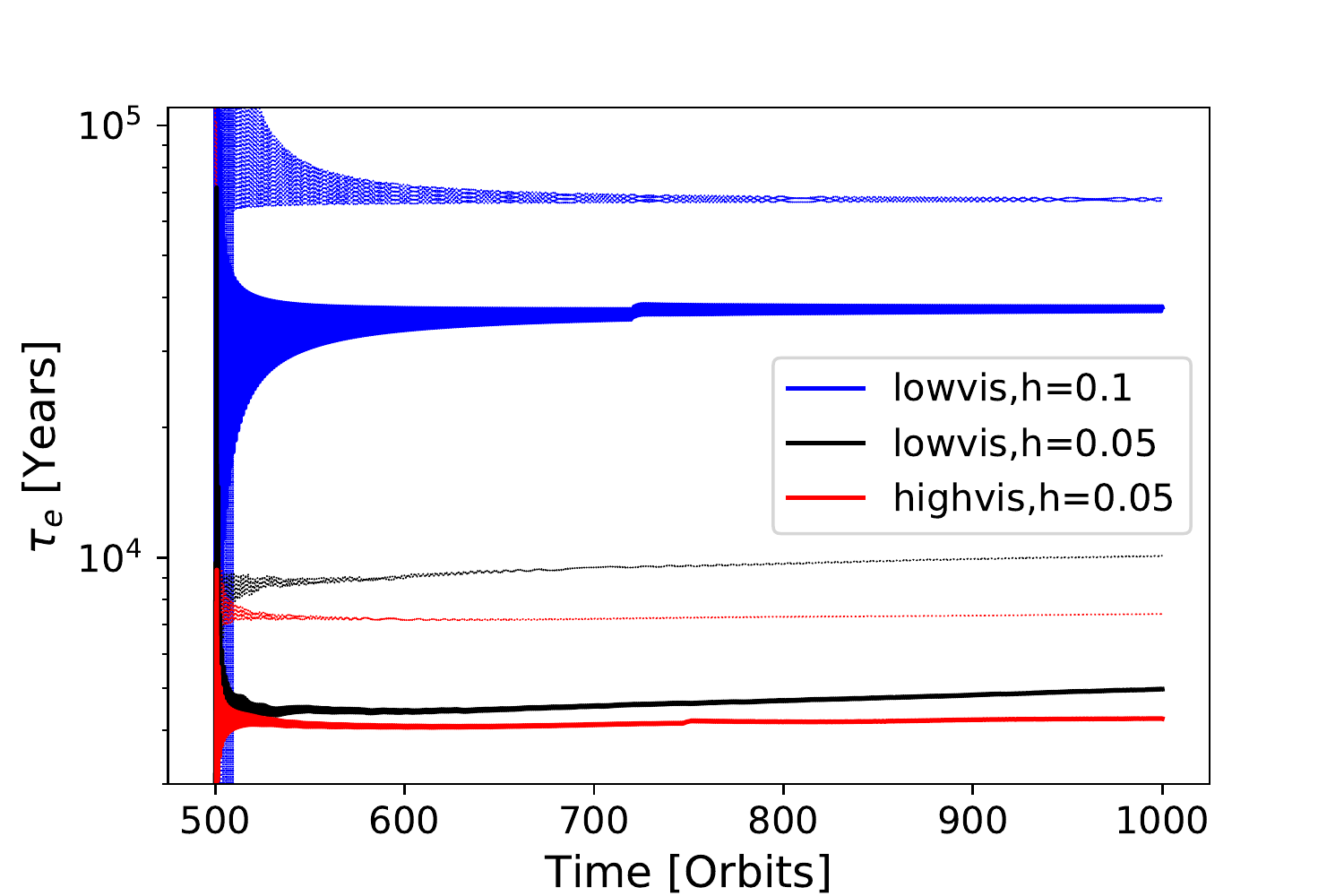}
  \caption{
%  \yxedit{
Running-time average of $\tau_e$ for $q=10^{-4}$ simulations. We start the averaging from 500 orbits to avoid uncertainties. The results from high viscosity $\alpha=10^{-3}$ and low viscosity $\alpha=10^{-4}$ simulations for $q=10^{-4}$ are shown in the red and black lines. Solid lines are from the fiducial cases with $\Sigma_{g,0}=8.9 \text{\ g\ cm}^{-2}, a_p=10$au and faint dotted lines are for additional cases with $\Sigma_{g,0}=3\text{\ g\ cm}^{-2}, a_p=30$au. The reason for low viscosity cases to have larger eccentricity damping timescale is due to the ability of the planet to open a partial gap and reduce $\Sigma_g$ in the proximity. Two cases with $\alpha=10^{-4}$ and $h_0=0.1$ are shown in blue lines.}
%}
  \label{taue}
\end{figure}

In our simulations, we have fixed the planets' semi-major axis and eccentricities for simplicity. 
However, it's crucial 
to investigate if such eccentricity could be maintained if the planet was released. Although we do not explicitly compute 
the disk's feedback on the dynamical evolution of the planet, 
its damping efficiency can be evaluated  
in terms of the power $\mathcal{P}$ and torque $\mathcal{T}$ exerted on the planet by the disk gas \citep[e.g.][]{DuffelChiang2015}: 

\begin{equation}
\left\langle\frac{\dot{e}}{e}\right\rangle=\frac{\langle \mathcal{P}\rangle\left(1-e^{2}\right)-\Omega_{0} \langle \mathcal{T}\rangle \sqrt{1-e^{2}}}{\Omega_{0}^{2} a^{2} M_{\mathrm{p}} e^{2}},
\end{equation}
where we use $\langle X\rangle$ to denote the running-time average of quantity $X$ from a certain initial time until the current 
orbit, and define the average damping timescale $\tau_e$ as $-1/\langle{\dot{e}}/{e}\rangle$. 
To avoid uncertainties, we start averaging from 500 orbits, and the result of $\tau_e$ in $q=1\mathrm{e}-4$ cases are plotted in Fig \ref{taue} for the fiducial as well as additional simulations. We find
that for $q=10^{-4}$, the gas damping timescale $\tau_e$ for $e=0.1, h_0=0.05$ converges to a few hundred orbits without external forcing, in agreement with the findings of \citet{Cresswell2007,BitschKley2010,Li2019}, equivalent to a few to ten thousand years in physical units for $a_p=10-30$au. These are relatively small fractions of the disk lifetime. On the other hand, \citet{bailey} applied high resolution 3D simulations and found that with small smoothing length, eccentric planets are able to significantly perturb the flow field in its vicinity and reduce the damping torque to be much smaller than linear predictions, but this cannot be directly verified with our large smoothing length simulations.

A potential channel of prolonging the damping timescale and making such features more common is to have a larger aspect ratio $h_0$ at planet location \citep{Artymowicz1993,DuffelChiang2015,Li2019}. Indeed a larger $h_0$ is possible for passively irradiated disks at large separations $\geq 20-30$au \citep{GaraudLin2007, Chiang2013}. For example, we have run two extra low viscosity simulations with $h_0=0.1$, in which case $\tau_e$ converges to much larger 
%\yxedit{
values than in the $h_0=0.05$ cases, as shown in Fig \ref{taue}. However, we found that this configuration does not have such evident gap features in the $\mathrm{St} \sim 0.015-0.04$ dust profile as the $h_0=0.05$ cases. Nevertheless, at even larger distances $a_p\gtrsim 50$au, a lower gas surface density that leads to larger Stokes number of the dust may help to manifest an observable dust-gap feature, the possibility of which is prone to further studies.
%}

We also note that our numerical models are computed for laminar disk flows.  In realistic PPDs, 
turbulence can lead to random torques
\citep{nelsonpap2003, johnsongoodman2006, baruteaulin2010} and low-level eccentricity excitation \citep{nelson2005}.
Tidal torque induced by planets with circular orbits on disks with weak turbulence (i.e. low $\alpha$) may
also leads to RWI-induced vortices in the disk and low-level eccentricity for the planets.  

Another channel to induce and maintain eccentricity is through long-term perturbation, a common phenomenon among progenitors of ubiquitously observed multiple super-Earth or Neptune companions. 
%\yxedit{
Indeed among the 39 planetary gap samples utilized by \citet[][see their Table A1]{Nayakshin2019}, 26 samples are taken from 10 systems harboring double or triple gap features, implicitly assuming co-existence of multiple planets in these PPD systems.
%} 
Before they gain enough mass to open a gap, these protoplanets undergo type I migration, whose pace and direction is determined, among other things, by the planets mass, and the profiles of the gas surface density and temperature around the planet’s orbital radius. \citep{Paardekooperetal2010a, Paardekooperetal2011}.  In typical PPDs, embedded protoplanets'
migration tends to converge to the characteristic boundary between inner region heated by viscous dissipation and 
outer region heated by irradation, the dead zone boundaries, or snow lines  \citep{Kretke_Lin_2012, 
Bitsch2015, liu2015,  liu2016} where disk opacity changes, as well as the radii of maximum viscous stress \citep{LydenBellPringle1974,   chen2020b} far from the host star.  Convergent migration naturally leads protoplanets to capture each other 
into their mutual resonances \citep{Zhang2014, liu2015}.  Through combined resonant interaction between planets and their natal disks, planet pairs may maintain eccentricities at a few $10^{-2}$ level \citep{Zhang2014}, in contrast to the damping tendency for 
single isolated planets (Fig \ref{taue}). An advantage of this mechanism is while isolated planets with small eccentricities could smear out double-ring structures - commonly observed in ALMA samples \citep{Huang2018} and attributed to single planets with circular orbits in low-viscosity disks \citep[also see \S\ref{sec:lowvis}]{Dong2017, Dong2018} - planet pairs at resonance locations might recover this feature.

%The combined torque imposed by planets on the disk also leads to the formation of multiple gaps separated in distance, comparable to that simulated with . 

In relatively massive disks, the feedback torque from the disks dominates over the interaction between 
super-Earths, enables them to bypass their mutual resonant barriers, merge, form super-critical
cores, undergo runaway gas accretion, and to form one or more gas giants \citep{liu2015, liu2016, Ida2013}.
Under influence of type II migration, Jupiter mass planets also capture each other into stationary or mean motion resonances \citep{leepeale2002, 
kleylee2005}, and may open a common gap in both the gas and the dust profiles \citep{Brydenetal2000, dongzhu2015,  
duffelldong2015,Marzari2019}. In contrast, much less massive super-Earth/Neptune pairs still keep shallow but 
independent gaps in the gas profile \citep{Papaloizou2005, Zhang2014, Kanagawa2020}, albeit its effect on observable dust 
signatures in PPDs remains to be probed by hydrodynamical simulations. The detail of this scenario, as well as 
the specific mass threshold between planet pairs that open up a combined cavity or individual gaps in the dust profiles 
remains to be studied in subsequent works,  
%\yxedit{
which also serves as a test of the assumption that multiple gaps in the \citet{lodato2019,Nayakshin2019} samples may  indeed be used to infer the mass of multiple planets.
%}
%We calculated the gap-ring distance $w'$ in the dust density distribution, instead of the 

\section{Summary}
\label{summary}

In this paper, we performed hydrodynamical simulations and Monte Carlo analysis to show that planets' eccentricity brings a degeneracy in the method of mass inference from the gap widths/gap-ring separations. We emphasize on the role of eccentricity in expanding the dust gap when the radial excursion is significantly larger than its Hill radius, and that considering only the empirical scaling for circular planets may introduce artificial discrepancies between the retrieved mass population and predictions from basic core accretion theory, smoothing out a deficit in the abundance of sub-Saturns in the midst of runaway accretion. 
% -overpredicts Jupiters? can be explained by \citep{lichen2021} reduce the dynamical cross section efficiently. 
Moreover, we also note that eccentric planets induce more symmetric dust profiles than circular planets in low viscosity environments, which might reconcile the general low viscosity in PPDs with the lack of observed symmetries in ALMA observations. 

We have applied 2D simulations to model partial gap opening and pebble-isolation of super Earths and Neptunes, which is a 3D process \citep{Bitsch2018}. But this circumstance might not significantly affect the big grains settled onto the midplane - the location where ALMA observations directly probe - and their radial dynamics. 
%\yxedit{
Nevertheless, for dust with St $\approx 0.01-0.03$, our simulations are run 
for less than a radial drift time across the entire computational domain and the results therefore represent an temporary response to perturbation of the local gas density rather than a secular steady state. The time-dependence of the ring locations remains to be further scrutinized.
%}

To extend the general conclusion in this study, there are multiple directions for future investigation. A direct extension is to apply more hydrodynamical simulations to constrain a more rigorous scaling (e.g. fit the constants $A, B$ in Eqn \ref{wrformula}) of the characterstic gap-ring distance with respect to planet mass and eccentricity, under a variety of $\alpha, h_0$ disk parameters. We can also explore possible observation signatures for identifying eccentric planets in a gap-ring system and break the mass-eccentricity degeneracy with synthetic images. 

{A notable difference between raw density distributions of dust and the sub-mm images is the location of gap center. In our work we have used a simple prescription to define $w^\prime$ and $r_{gap}$, and only in the $q=10^{-4}, \alpha=10^{-3}$ eccentric case $r_{gap}$ is estimated to be considerably smaller than $a_p$, which strongly increases normalized $w^\prime/r_{gap}$ despite an insignificant retreat of pressure maxima. 
However, it was implied in \citet{rosotti2016} that planetary gaps in sub-mm dust images often extend deep into the inner disk (the gaps nearly become cavities), such that an underestimated $r_{gap}$ is very common even for planets on circular orbits. This outstanding issue led them to use scattered-light images to define an $r_{gap}$ which represents the planet's orbital radius more closely (see their Fig 12 \& 13). On the other hand, scattered-light images are not applied in gap-width samples in the DSHARP survey \citep[e.g.][]{long2018,andrews2018,Long2019}. The consistency between these different methods remains to be investigated.}

Subsequent studies may include consistent dust feedback and coagulation \citep[e.g.][]{li2019coag,Li2020,Laune2020} to investigate how more realistic conditions affect the picture. Although \citet{fu2014} showed that including dust feedback might damp the RWI and reduce vortex lifetimes, the problem of asymmetry maintenance is not the main focus of our paper and we only point out the damping of asymmetry in low viscosity environments as one side-effect of planet eccentricity. Whether feedback effects would significantly affect the azimuthally averaged dusty profiles (especially the radial location of the peaks) is a separate problem deserving attention on its own right.

A parallel direction for future research is the possible origin of long-term eccentricity for
isolated or multiple closely-packed planets embedded in young disks. 
At large distances to the host star, an aspect ratio of $h_0\gtrsim 0.1$ might maintain the planet eccentricity 
for a considerable fraction of disk lifetime, in which case dust with larger Stokes number is 
needed to produce an observable gap-ring structure. Random torques due to turbulent flows in the PPD may also induce eccentricity of planets. Another novel mechanism for eccentricity maintenance is for 
dust-gap-opening planets to have companions which provide secular resonant perturbations. This possibility is particularly relevant as 
the progenitors of multiple super Earths are commonly found in exoplanet surveys, 
%\yxedit{
and consistent with the assumptions in \citet{lodato2019,Nayakshin2019} that multiple gaps in one PPD system could indicate multiple planets.
%} 
Such configuration may be a natural consequence of sequential planet formation and migration 
scenarios, albeit whether the planet pair would open a 
combined large cavity or individual gaps in the dust profile remains to be investigated.

%The low viscosity inference is consistent with numerical simulations of magnetorotational instabilities in PPDs \citep{bai2013}. 

\acknowledgements
Y-X. C thanks Chris Ormel, Haochang Jiang and Shuo Huang for helpful discussions. We thank Xue-Ning Bai and Yi Mao for providing computational resources. We thank the anonymous referee for helpful comments which improved this paper. The LANL Report number is LA-UR-21-28245.

\bibliography{ref_global}
\bibliographystyle{aasjournal}

% the difference between these power law scaling may be small, we propose that eccentricity will lead to misleading results. In the next section, we apply numerical simulation to demonstrate out point. In particular.

\end{document}